\documentclass[12pt]{article}
\usepackage{epsfig}
\usepackage{hyperref}%

\date{}
\begin{document}

\newcommand{\beq}{\begin{equation}}
\newcommand{\eeq}{\end{equation}}
\newcommand{\nn}{\nonumber}
\newcommand{\bea}{\begin{eqnarray}}
\newcommand{\eea}{\end{eqnarray}}

\title{Brane World Dynamics and Conformal Bulk Fields}

\author{Rui Neves\footnote{E-mail: \tt rneves@ualg.pt}\hspace{0.2cm}
  and
Cenalo Vaz\footnote{E-mail: \tt cvaz@ualg.pt}\\
{\small \em Faculdade de Ci\^encias e Tecnologia,
Universidade do Algarve}\\
{\small \em Campus de Gambelas, 8000-117 Faro, Portugal}
}

\maketitle

\begin{abstract}
In the Randall-Sundrum scenario we investigate the dynamics of
a spherically symmetric 3-brane world
when matter fields are present in the bulk. To analyze the
5-dimensional Einstein equations we employ a
global conformal transformation whose factor characterizes
the $Z_2$ symmetric warp. We find a new set of exact dynamical
collapse solutions which localize gravity in the
vicinity of the brane for a stress-energy tensor
of conformal weight -4 and a warp factor that depends only on the
coordinate of the fifth dimension. Geometries which describe the dynamics of
inhomogeneous dust and generalized dark radiation on the brane
are shown to belong to this set. The conditions for singular or
globally regular behavior and the static marginally bound limits are
discussed for these examples. Also explicitly demonstrated is complete
consistency with the effective point of view of a 4-dimensional
observer who is confined to the brane and makes the same assumptions
about the bulk degrees of freedom.
\vspace{0.5cm}

\end{abstract}

\section{Introduction}

Matter fields may be confined to a 3-brane world embedded in a
higher dimensional space if gravity propagates away from the brane
in the extra dimensions \cite{EBWM,CED}. Remarkably, in this
context it is possible to reformulate the hierarchy problem
following two alternative paths which admit a fundamental Planck
scale in the TeV range. In one approach the extra dimensions need
to be compactified to a large finite volume \cite{CLED}. In the
other, the so-called RS1 model, one considers two branes, one of
which (the observable brane) must have a negative tension. An
exponential hierarchy is generated by a ``warp'' factor that
characterizes the background metric \cite{RS1}-\cite{TM}. An
alternative setup involving infinite, non-compact extra dimensions
and only positive tension branes was later proposed in \cite{LR}.

It is also possible to consider a single brane together with infinite
and non-compact extra dimensions (the so-called RS2 model). The warp
localizes gravity in the vicinity of the brane, allowing for the
recovery of 4-dimensional Einstein gravity at low energy scales
\cite{RS2}-\cite{GiR}. In this framework the observable universe is a
boundary hypersurface of an infinitely extended $Z_2$ symmetric 5-dimensional
anti-de Sitter (AdS) space. The classical dynamics are defined by the
Einstein equations with a negative bulk cosmological constant $\Lambda_B$,
a Dirac delta source representing the brane and a stress-energy tensor
describing other bulk field modes which may exist in the whole AdS space.

A set of vaccum solutions is given by the metric \cite{CHR}
\beq
d{\tilde{s}_5^2}={{l^2}\over{{\left(|z-{z_0}|+{z_0}\right)}^2}}
\left(d{z^2}+d{s_4^2}\right),
\eeq
where $d{s_4^2}$ is a 4-dimensional line element generated by a Ricci
flat metric. The parameter $l$ is the AdS radius and is written as
$l=1/\sqrt{-{\Lambda_B}{\kappa_5^2}/6}$ where ${\kappa_5^2}=8\pi/{M_5^3}$ with
$M_5$ the fundamental 5-dimensional Planck mass. The brane is located
at $z={z_0}$ and is fine-tuned to have zero cosmological constant,
giving ${\Lambda_B}=-{\kappa_5^2}{\lambda^2}/6$ or $l=6/({\kappa_5^2}\lambda)$
where $\lambda$ denotes the positive brane tension.

The original RS solution \cite{RS2} belongs to this class and is
obtained when the 4-dimensional subspace has a Minkowski
metric. Another example is the black string solution \cite{CHR} which
induces the Schwarzschild metric on the brane \footnote{This solution was first
discussed in a different context in R. C. Myers and M. J. Perry, Ann. Phys.
{\bf 172}, 304 (1986).}. However, the Kretschmann scalar diverges both
at the AdS
horizon and at the black string singularity \cite{CHR} leading to a violation
of the cosmic censorship conjecture \cite{CCC}. This solution is
expected to be unstable
near the AdS horizon \cite{GLinst,Glinst1} and is therefore not an acceptable
description of a black hole on the brane. It may decay to a black
cylinder localized near the brane which is free from naked
singularities \cite{CHR} but, despite considerable effort, this
still remains a conjecture.

While exact solutions interpreted as static black holes localized on a
brane have been found for a 2-brane embedded in a 4-dimensional AdS space
\cite{EHM}, a static
black hole localized on a 3-brane is yet to be discovered
\cite{SS}-\cite{KOT}. The difficulty in finding a static solution has
led to an interesting conjecture \cite{EFK} which attempts to relate
black hole solutions localized on the brane in an AdS${}_{D+1}$
braneworld, which are found with the brane boundary conditions, with
quantum black holes in $D$ dimensions rather than classical ones.

Many other 5-dimensional solutions have been determined within the RS
brane world scenario. These include the extensions of the RS geometry
to thick branes \cite{CEHS} and non-fine-tuned branes \cite{KR}, the
5-dimensional metrics describing Friedmann-Robertson-Walker (FRW) cosmologies
\cite{5DBWCos,CosDR} and also other solutions envolving
the effect of scalar fields in the bulk \cite{SBSF}-\cite{CGRT}. The
aplication of the covariant Gauss-Codazzi (GC) approach
\cite{CGC}-\cite{RM} describing the perspective of a 4-dimensional
observer restricted to the brane, has permitted the analysis of even
more brane world physics \cite{DMPR}-\cite{RM4dp}. In addition,
several connections have been established between the AdS/CFT
correspondence \cite{CFT} and the RS brane world scenario \cite{GKR,EFK,RSCFT}.

In spite of the numerous 5-dimensional exact solutions already found
in the RS scenario there are still many effective 4-dimensional
metrics which have not been shown to be associated with exact bulk
spacetimes. This happens for instance with the inhomogeneous dust and
dark radiation dynamics on the brane. While the latter has been
deduced within the RS scenario using the covariant GC formulation
\cite{RC}, the former has not yet been shown to have a brane world
description. It is thus the purpose of the present work to search for
new exact 5-dimensional solutions which might reproduce such effective
4-dimensional brane world dynamics. To attain this objective we
consider the generally inhomogeneous dynamics of a spherically
symmetric RS 3-brane when field modes other than the Einstein-Hilbert
gravity are present in the whole AdS space. Such bulk fields and their
influence on the brane world dynamics have been extensively discussed
within the RS scenario for example to stabilize the extra
dimensions \cite{GW,TM,SBSF,KKOP} and to analyze the possibility of
generating a black hole localized on a 3-brane world \cite{LP,KT}.

We begin in section 2 with an analysis of the Einstein field equations
for the most general 5-dimensional metric
consistent with the $Z_2$ symmetry and with spherical symmetry on the
brane. We organize the Einstein equations using a
global conformal transformation whose factor characterizes the warp of the
fifth dimension. If the stress-energy tensor is assumed to have conformal weight
$s=-4$, then it is possible to decouple the gravitational dynamics of the bulk
matter fields from the 5-dimensional warp. The bulk matter dynamics generate a non-zero
pressure along the fifth dimension, which must satisfy a precise
equation of state. Thus we obtain a new set of exact dynamical solutions
for which gravity is localized near the brane by the conformal warp factor.

In sections 3 and 4 we analyze two examples which belong to this new
set of exact 5-dimensional solutions. Both examples permit the use of
synchronous coordinates. From the point of view of an observer confined
to the brane, they correspond to bulk matter behaving on the brane as
pressureless dust and a generalized form of dark radiation. We determine the static
marginally bound limits and discuss the conditions for singular or globally
regular behavior.

As a consistency check, we consider the point of view of an observer
confined to the brane in section 5. Using the effective GC
formulation we show that if such an observer uses the same description
of the field variables, then it indeed leads to identical brane world
dynamics. Explicit proofs are provided for dust and dark radiation.
We conclude in section 6.

\section{5-Dimensional Einstein Equations and Conformal
Transformations}

Let $(t,r,\theta,\phi,z)$ be a set of comoving coordinates in the 5-dimensional bulk.
The most general metric consistent with the $Z_2$ symmetry in $z$ and with 4-dimensional
spherical symmetry on the brane may be written using a global conformal transformation \cite{WaldBD}
defined by a $Z_2$ symmetric warp function $\Omega=\Omega(t,r,z)$,
\beq
{\tilde{g}_{\mu\nu}}={\Omega^2}{g_{\mu\nu}}.\label{gm4}
\eeq
The corresponding 5-dimensional line elements are related by
\beq
d{\tilde{s}_5^2}={\Omega^2}d{s_5^2},
\eeq
where
\beq
d{s_5^2}=d{z^2}+d{s_4^2}\label{gm2}
\eeq
has a 4-dimensional line element which depends on three $Z_2$ symmetric functions $A=A(t,r,z)$, $B=B(t,r,z)$ and
$R=R(t,r,z)$ as follows
\beq
d{s_4^2}=-{e^{2A}}d{t^2}+{e^{2B}}d{r^2}+{R^2}d{\Omega_2^2}.\label{gm3}
\eeq
$R(t,r,z)$ represents the physical radius of the 2-spheres.

When bulk field modes other than the Einstein-Hilbert gravity are present in the 5-dimensional space the dynamical
RS action corresponding to $\tilde{g}_{\mu\nu}$ is given by
\beq
\tilde{\mathcal{S}}=\int{d^4}xdz\sqrt{-\tilde{g}}
\left[{\tilde{R}\over{2{\kappa_5^2}}}-
{\Lambda_B}-{\lambda\over{\sqrt{\tilde{g}_{55}}}}\delta\left(z-{z_0}\right)+
{\tilde{\mathcal{L}}_B}\right].\label{5Dact}
\eeq
The brane is assumed to be located at $z={z_0}$ and is a fixed point of the $Z_2$ symmetry of the manifold. The
contribution of the bulk fields is defined by the lagrangian $\tilde{\mathcal{L}}_B$. A Noether variation on the
action (\ref{5Dact}) gives the classical Einstein field equations,
\beq
{\tilde{G}_\mu^\nu}=-{\kappa_5^2}\left[{\Lambda_B}{\delta_\mu^\nu}+
\lambda\delta
\left(z-{z_0}\right){\tilde{\gamma}_\mu^\nu}-{\tilde{T}_\mu^\nu}\right],
\label{5DEeq}
\eeq
where the induced metric on the brane is
\beq
{\tilde{\gamma}_\mu^\nu}={1\over{\sqrt{\tilde{g}_{55}}}}\left(
{\delta_\mu^\nu}-{\delta_5^\nu}{\delta_\mu^5}\right)
\eeq
and the stress-energy tensor associated with the bulk fields is defined by
\beq
{\tilde{T}_\mu^\nu}={\tilde{\mathcal{L}}_B}{\delta_\mu^\nu}-
2{{\delta{\tilde{\mathcal{L}}_B}}\over{\delta{\tilde{g}^{\mu\alpha}}}}
{\tilde{g}^{\alpha\nu}}
\eeq
and is conserved in the bulk,
\beq
{\tilde{\nabla}_\mu}{\tilde{T}_\nu^\mu}=0.\label{5Dceq}
\eeq

The Einstein field equations (\ref{5DEeq}) are extremely complex when the metric $\tilde{g}_{\mu\nu}$ is
considered in its full generality. To be able to solve them we need to introduce simplifying assumptions
about the field variables involved in the problem. Let us first assume that under the conformal transformation
(\ref{gm4}) the bulk stress-energy tensor has conformal weight\footnote{Our assignment of conformal weight
depends upon the index position.} $s$ \cite{WaldBD},
\beq
{\tilde{T}}^{\mu\nu}={\Omega^{s}}T^{\mu\nu}.\label{ctst}
\eeq
Then substituting Eq. (\ref{ctst}) and using the known transformation properties of the other tensors
\cite{WaldBD} we re-write Eq. (\ref{5DEeq}) as
\bea
{G_\mu^\nu}&=&-6{\Omega^{-2}}\left({\nabla_\mu}\Omega\right){g^{\nu\rho}}
{\nabla_\rho}\Omega+
3{\Omega^{-1}}{g^{\nu\rho}}{\nabla_\rho}{\nabla_\mu}\Omega
-3{\Omega^{-1}}{\delta_\mu^\nu}{g^{\rho\sigma}}{\nabla_\rho}{\nabla_\sigma}
\Omega\nn\\
&&-{\kappa_5^2}
{\Omega^2}\left[{\Lambda_B}{\delta_\mu^\nu}+\lambda{\Omega^{-1}}
\delta(z-{z_0}){\gamma_\mu^\nu}-{\Omega^{s+2}}{T_\mu^\nu}\right].
\label{t5DEeq}
\eea
Similarly, Eq. (\ref{5Dceq}) also transforms under the conformal transformation. We have
\beq
{\nabla_\mu}{T_\nu^\mu}+{\Omega^{-1}}\left[(s+7){T_\nu^\mu}{\partial_\mu}
\Omega-T{\partial_\nu}\Omega\right]=0,\label{t5Dceq}
\eeq
where $T={T_\mu^\mu}$ is the trace of the bulk stress-energy tensor. If $\tilde{T}_\mu^\nu$ has conformal weight
$s=-4$, it is possible (though not necessary) to separate
Eq. (\ref{t5DEeq}) as follows
\beq
{G_\mu^\nu}={\kappa_5^2}{T_\mu^\nu},\label{r5DEeq}
\eeq
\bea
&6{\Omega^{-2}}{\nabla_\mu}\Omega{\nabla_\rho}
\Omega{g^{\rho\nu}}-
3{\Omega^{-1}}{\nabla_\mu}{\nabla_\rho}\Omega{g^{\rho\nu}}+3{\Omega^{-1}}
{\nabla_\rho}{\nabla_\sigma}\Omega{g^{\rho\sigma}}{\delta_\mu^\nu}=\nn\\
&-{\kappa_5^2}
{\Omega^2}\left[{\Lambda_B}{\delta_\mu^\nu}+\lambda{\Omega^{-1}}
\delta(z-{z_0}){\gamma_\mu^\nu}\right].\label{5DEeqwf}
\eea
In Eq. (\ref{r5DEeq}) we find the 5-dimensional Einstein equations with fields present in the bulk when the
brane and the bulk cosmological constant are absent. It does not depend on the conformal warp factor which is
dynamically defined by Eq. (\ref{5DEeqwf}) and is then the only effect reflecting the existence of the brane
or of the bulk cosmological constant in this setting. We emphasize that this is only possible for the special
class of bulk fields which have a stress-energy tensor with conformal weight $s=-4$. The decomposition
of (\ref{5DEeq}) according to (\ref{r5DEeq}) and (\ref{5DEeqwf}) implies a separation of Eq. (\ref{t5Dceq})
because of the Bianchi identity. We must have
\beq
{\nabla_\mu}{T_\nu^\mu}=0,\label{r5Dceq}
\eeq
\beq
3{T_\nu^\mu}{\partial_\mu}\Omega-T {\partial_\nu}\Omega=0.
\label{5Dceqwf}
\eeq
Note that now $T_\mu^\nu$ is a conserved tensor field which must satisfy the additional warp constraint equations
(\ref{5Dceqwf}). Since $s\not=-7$ it does not need to be traceless.

Let us now consider Eq. (\ref{r5DEeq}). Expanding the Einstein tensor in terms of the metric functions $A$, $B$
and $R$ reveals that its only non-zero off-diagonal elements are $G_t^r$, $G_t^z$ and $G_r^z$. Assuming that
$A=A(t,r)$, $B=B(t,r)$ and $R=R(t,r)$ sets $G_t^z$, $G_r^z$ and the corresponding stress-energy tensor components
to zero. Hence, we have
\beq
{T_a^z}=0,\label{eqst1}
\eeq
where the latin index represents the 4-dimensional brane coordinates $t$, $r$, $\theta$ and $\phi$.
If in addition $\Omega=\Omega(z)$ then Eq. (\ref{5DEeqwf}) turns out to be independent of the metric functions
$A$, $B$, $R$ and reads
\[
6{\Omega^{-2}}{{({\partial_z}\Omega)}^2}=
-{\kappa_5^2}{\Omega^2}{\Lambda_B},
\]
\beq
3{\Omega^{-1}}{\partial_z^2}\Omega=-{\kappa_5^2}{\Omega^2}
\left[{\Lambda_B}+\lambda{\Omega^{-1}}\delta(z-{z_0})\right].\label{rswf}
\eeq
For definiteness, take as solution of Eq. (\ref{rswf}) the RS conformal warp factor
\beq
\Omega={\Omega_{\mbox{\tiny RS}}}\equiv {l\over{|z-{z_0}|+{z_0}}}\label{RSwf1}
\eeq
where $l=\sqrt{-6/({\Lambda_B}{\kappa_5^2})}$ and
\beq
{\Lambda_B}+{{{\kappa_5^2}{\lambda^2}}\over{6}}=0.\label{RSwf2}
\eeq
Of course, other solutions with warp factors which depend only on the
5-dimensional coordinate $z$ such as those corresponding to
non-fine-tuned branes \cite{KR} or thick branes \cite{CEHS} may also
be considered (see \cite{KT}). Eq. (\ref{5Dceqwf}) constrains
$T_\mu^\nu$ to satisfy the equation of state
\beq
2{T_z^z}={T_c^c}.\label{eqst2}
\eeq
If we consider a diagonal stress-tensor,
\beq
{T_\mu^\nu}=diag\left(-\rho,{p_r},{p_T},{p_T},{p_z}\right),\label{bmten}
\eeq
where $\rho$, $p_r$, $p_T$ and $p_z$ denote the bulk matter density and
pressures, then Eq. (\ref{eqst2}) is
re-written as
\beq
\rho-{p_r}-2{p_T}+2{p_z}=0.\label{eqst3}
\eeq
Because the metric functions $A$, $B$ and $R$ do not depend on $z$ Eq.
(\ref{eqst1}) applied to Eq. (\ref{r5Dceq}) leads to ${\partial_z}{p_z}=0$
and to
\beq
{\nabla_a}{T_b^a}=0,\label{4Dceq}
\eeq
Therefore $\rho$, $p_r$ and $p_T$ must also be independent of $z$. Matter is,
however, inhomogeneously distributed along the fifth dimension. The physical energy
density, ${\tilde \rho}(t,r,z)$, and pressures, ${\tilde p}(t,r,z)$, are related
to their counterparts, $\rho(t,r)$ and $p(t,r)$, by the scale factor $\Omega^{-2}(z)$.

Since all the off-diagonal components are zero, Eq. (\ref{r5DEeq})
also admits a similar dimensional reduction
\beq
{G_a^b}={\kappa_5^2}{T_a^b}\label{4DEeq}
\eeq
but now $G^z_z$ is in general non-zero because of the existing pressure $p_z$,
\beq
{G_z^z}={\kappa_5^2}{p_z}.\label{5DEeqz}
\eeq
The collapse of the conformal bulk matter is inhomogeneous and defined by Eqs. (\ref{4Dceq})
and (\ref{4DEeq}). As we show in section 5, gravity is localized in the vicinity of the brane.
The matter dynamics  generates a pressure $p_z$ along the fifth dimension, which must
consistently be given by Eqs. (\ref{eqst3}) and (\ref{5DEeqz}). There are no further
constraints.

\section{Dust Dynamics on the Brane}

The simplest kind of matter which we may consider is characterized by the
equation of state ${p_r}={p_T}=0$. According to Eq. (\ref{eqst3}), its density
$\rho={\rho_{\mbox{\tiny D}}}$ must
generate a pressure $p_z$ along the fifth
dimension which satisfies ${p_z}=-{\rho_{\mbox{\tiny D}}}/2$. One can mimic an
effective brane cosmological constant, if instead we take
\beq
\rho={\rho_{\mbox{\tiny D}}}+{\Lambda\over{\kappa_5^2}},\quad {p_r}={p_T}=
-{\Lambda\over{\kappa_5^2}},\label{dustes1}
\eeq
giving
\beq
{p_z}=-{1\over{2}}\left({\rho_{\mbox{\tiny
        D}}}+4{\Lambda\over{\kappa_5^2}}
\right).\label{dustes2}
\eeq
Although we will shortly see that $\Lambda$ mimics a cosmological constant
on the brane, it should be emphasized that $\Lambda$ is a bulk quantity, distinct from
a 4-dimensional cosmological constant. Since the warp factor has already been chosen to be the RS
solution (\ref{RSwf1}), in order to find the 5-dimensional metric describing the dynamics
we must solve Eqs. (\ref{4Dceq}), (\ref{4DEeq}) and (\ref{5DEeqz})
under conditions (\ref{dustes1}) and (\ref{dustes2}).

Using Eq. (\ref{dustes1}) we write Eq. (\ref{4Dceq}) \cite{TPS} as follows
\beq
\dot{B}{\rho_{\mbox{\tiny D}}}=-\dot{\rho_{\mbox{\tiny
      D}}}-2{\dot{R}\over{R}}
{\rho_{\mbox{\tiny D}}},\label{dceq1}
\eeq
\beq
A'{\rho_{\mbox{\tiny D}}}=0,\label{dceq2}
\eeq
where the dot and the prime denote, respectively, partial
differentiation with respect to $t$ and $r$. Because of Eq. (\ref{dceq2})
we have to take the synchronous frame where $A=0$ to avoid setting
$\rho_{\mbox{\tiny D}}$ to zero. Then the off-diagonal equation is given by
\beq
{G_r^t}={2\over{R}}\left(\dot{R}'-\dot{B}R'\right)=0
\eeq
and has the solution
\beq
{e^B}={R'\over{H}},\label{eB}
\eeq
where $H=H(r)$ is an arbitrary positive function of $r$. Introducing
Eq. (\ref{eB}) in Eq. (\ref{dceq1}) we get the required dust density \cite{PSJ},
\beq
{\rho_{\mbox{\tiny D}}}={{2{G_N}M'}\over{{\kappa_5^2}{R^2}R'}},\label{dustd}
\eeq
where $G_N$ is Newton's gravitational constant and $M=M(r)$ is an arbitrary positive function of $r$ which represents the
dust mass inside a shell labelled by $r$. Note that ${\rho_{\mbox{\tiny D}}}>0$ is equivalent to $M'/R'>0$. This
implies that the weak, strong and dominant energy conditions
\cite{WaldBD} are satisfied in 4 and 5 dimensions.

Next, consider the trace equation
\beq
-{G_t^t}+{G_r^r}+2{G_\theta^\theta}=-2{{\ddot{R}'}\over{R'}}-
4{\ddot{R}\over{R}}={{2{G_N}M'}\over{{R^2}R'}}-2\Lambda.
\eeq
Integrating twice we find
\beq
{\dot{R}^2}={{2{G_N}M}\over{R}}+{\Lambda\over{3}}{R^2}+f,
\eeq
where $f\equiv f(r)$ is an arbitrary function of $r$ to be interpreted as the energy inside a shell labelled by
$r$. Imposing on the initial hypersurface $t=0$ the condition $R(0,r)=r$ we obtain
\beq
\pm t+\psi=\int{{dR}\over{\sqrt{{{2{G_N}M}\over{R}}
+{\Lambda\over{3}}{R^2}+f}}},\label{dust2}
\eeq
where the signs $+$ or $-$ refer to expansion or collapse and $\psi=\psi(r)$ is given by the evaluation at $t=0$
of the integral in the r.h.s. Applying the radial equation
\beq
{G_r^r}=-2{\ddot{R}\over{R}}+{{H^2}\over{R^2}}-{1\over{R^2}}-
{{\dot{R}^2}\over{R^2}}=-\Lambda
\eeq
we obtain $H=\sqrt{1+f}$ and this restricts $f$ to satisfy $f>-1$. Then it is easy to see that Eq. (\ref{5DEeqz}),
\beq
{G_z^z}=-{{\ddot{R}'}\over{R'}}-2{\ddot{R}\over{R}}+
{{f-{\dot{R}^2}}\over{R^2}}
+{{(f-{\dot{R}^2})'}\over{RR'}}=-{{{G_N}M'}\over{{R^2}R'}}-2\Lambda,
\eeq
is an identity for all $M$ and $f$.

We have thus obtained the 5-dimensional dust collapse solutions
\beq
d{\tilde{s}_5^2}={\Omega_{\mbox{\tiny RS}}^2}\left(d{z^2}+d{s_4^2}
\right),\label{vacs1}
\eeq
where the 4-dimensional metric has the LeMa\^{\i}tre-Tolman-Bondi (LTB)\cite{LM,TB} form
\beq
d{s_4^2}=-d{t^2}+{{{R'}^2}\over{1+f}}d{r^2}+{R^2}d{\Omega_2^2},\label{vacs2}
\eeq
with the physical radius satisfying Eq. (\ref{dust2}).

The marginally bound models (corresponding to $f=0$) with constant mass function, $M(r) = \mathrm{const.}$, describe
static solutions. Indeed, using the standard transformation from the LTB coordinates $(t,r)$ to the curvature
coordinates $(T,R)$,
\beq
T=t+\int dR {{\sqrt{{\Lambda\over{3}}{R^4}+2{G_N}M R}}\over{
{\Lambda\over{3}}{R^3}-R+2{G_N}M}},
\eeq
we obtain an AdS/dS-Schwarzschild black string solution given by Eq. (\ref{vacs1}) where
\beq
d{s_4^2}=-\left(1-{{2{G_N}M}\over{R}}-{\Lambda\over{3}}{R^2}\right)d{T^2}+
{{\left(1-{{2{G_N}M}\over{R}}-{\Lambda\over{3}}{R^2}\right)}^{-1}}d{R^2}+
{R^2}d{\Omega_2^2}.
\eeq
In the vaccum we obtain the original RS static solution \cite{RS2} and not the Schwarzschild black string
\cite{CHR}. By Birkhoff's theorem, there are no other static vaccum solutions.

In general the solutions are dynamical and inhomogeneous. An
analysis of the potential $V={\Lambda\over{3}}{R^3}
+fR+2{G_N}M$ uncovers a rich set of singular and globally regular solutions \cite{DJCJ}.

\section{Generalized Dark Radiation Dynamics on the Brane}

Let us now consider the possibility of generating on the brane the localized gravitational interaction between
a generalized form of inhomogeneous dark radiation and an {\it effective} cosmological constant. This system is defined by
conformal bulk matter with the equations of state
\beq
\rho+{p_r}=0,\quad{p_T}+\eta\rho+{\Lambda\over{\kappa_5^2}}
\left(1-\eta\right)=0,\label{gdres1}
\eeq
where $\eta$ is the parameter characterizing the dark radiation model and $\rho$ is given by
\beq
\rho={\rho_{\mbox{\tiny DR}}}+{\Lambda\over{\kappa_5^2}}.\label{gdres2}
\eeq
Applying Eq. (\ref{eqst3}) we find
\beq
{p_z}=-\left(1+\eta\right)\rho-{\Lambda\over{\kappa_5^2}}\left(1-\eta\right).
\label{gdres3}
\eeq
As before, $\Lambda$ is a bulk quantity and not a 4-dimensional cosmological constant. Nevertheless it will mimic
a 4-dimensional cosmological constant on the brane. Thus only for standard dark radiation \cite{CosDR,RC} with $\eta=-1$
and $\Lambda=0$, is the fifth dimensional
pressure $p_z$ equal to zero. Furthermore, the trace of the stress-energy tensor is
\beq
{T_\mu^\mu}={T_a^a}+{p_z},
\eeq
and ${T_a^a}=2{p_z}$, so it only vanishes when $\eta=-1$ and
$\Lambda=0$. This implies that only the standard form
of dark radiation may be associated with the traceless projected Weyl tensor and so with the 4-dimensional brane
world vaccum in the effective Gauss-Codazzi approach \cite{RC}.

After decoupling the RS warp factor the determination of the 5-dimensional metric for the dark radiation system
requires the solution of Eqs. (\ref{4Dceq}), (\ref{4DEeq}) and (\ref{5DEeqz}) under conditions
(\ref{gdres1})-(\ref{gdres3}). Let us start by introducing Eqs. (\ref{gdres1}) and (\ref{gdres2}) in Eq. (\ref{4Dceq}).
The contribution of $\Lambda$ cancels out and we find the following dark radiation conservation equations
\beq
\dot{{\rho_{\mbox{\tiny DR}}}}+2(1-\eta){\dot{R}\over{R}}{\rho_{\mbox{\tiny
      DR}}}=0=
{\rho_{\mbox{\tiny DR}}}'+2(1-\eta){{R'}\over{R}}{\rho_{\mbox{\tiny
      DR}}}.\label{drceq1}
\eeq
Note that for this generalized dark radiation system we may also safely take the synchronous frame ($A=0$).
Because of the equation of state $\rho+{p_r}=0$ the dark radiation has a density defined independently of $A$
as a consequence of which, despite the existing pressures, it admits a synchronous solution. Note as well
that this is independent of the relation between $\rho$ and $p_T$. As a consequence the general equations
given in \cite{TPS} simplify to give Eq. (\ref{drceq1}). The corresponding inhomogeneous density solution
is given by

\beq
{\rho_{\mbox{\tiny
      DR}}}={{Q_\eta}\over{\kappa_5^2}}{R^{2\eta-2}}
\label{gdrden},
\eeq
where the constant $Q_\eta$ is the dark radiation tidal charge. If ${Q_\eta}>0$ then the dark radiation density
is positive. As a consequence the weak, the dominant and the strong
energy conditions in 4 dimensions imply,
respectively, $\eta\leq 1$, $|\eta|\leq 1$ and $\eta\leq 0$. If these conditions are imposed in 5 dimensions then
we find that they lead, respectively, to $\eta\leq 0$, $-2\leq\eta\leq 0$ and $\eta\leq -1/3$. Of course if
${Q_\eta}<0$ then the dark radiation density is negative and all the energy conditions are violated.

Substituting Eq. (\ref{gdrden}) in the Einstein trace equation we obtain
\beq
-{G_t^t}+{G_r^r}+2{G_\theta^\theta}=-2{{\ddot{R}'}\over{R'}}
-4{\ddot{R}\over{R}}=-2\eta{Q_\eta}{R^{2\eta-2}}-2\Lambda.
\eeq
After two integrations we find
\beq
{\dot{R}^2}={{Q_\eta}\over{2\eta+1}}{R^{2\eta}}+{\Lambda\over{3}}{R^2}
+f,\label{gdr1}
\eeq
where $f\equiv f(r)$ is the arbitrary function of $r$ interpreted as the energy inside a shell labelled by $r$.
Imposing the condition $R(0,r)=r$ on the initial hypersurface $t=0$ and integrating Eq. (\ref{gdr1}) we get
\beq
\pm t +\psi=\int{{dR}\over{\sqrt{{\Lambda\over{3}}{R^2}+f
+{{Q_\eta}\over{2\eta+1}}{R^{2\eta}}}}},\label{gdr2}
\eeq
Note that we have assumed $\eta\not=-1/2$. For $\eta=-1/2$ we obtain
\beq
{\dot{R}^2}={{Q_{\mbox{\tiny -1/2}}}\over{R}}\left(1+\ln R\right)
+{\Lambda\over{3}}{R^2}+f,\label{gdr11}
\eeq
and then
\beq
\pm t +\psi=\int{{dR}\over{\sqrt{{\Lambda\over{3}}{R^2}+f
+{{Q_{\mbox{\tiny -1/2}}}\over{R}}\left(1+\ln R\right)}}}.\label{gdr22}
\eeq
Note as well that if the evolution is to be dominated by $\Lambda$ as $R$ goes to infinity
then $\eta$ should satisfy $\eta<1$. This is true when ${Q_\eta}>0$ and any one of the energy conditions is
satisfied. Applying the radial equation
\beq
{G_r^r}=-2{\ddot{R}\over{R}}+{{H^2}\over{R^2}}-{1\over{R^2}}-
{{\dot{R}^2}\over{R^2}}=-{Q_\eta}{R^{2\eta-2}}-\Lambda
\eeq
we again obtain $H=\sqrt{1+f}$ with $f>-1$. Then Eq. (\ref{5DEeqz}),
\beq
{G_z^z}=-{{\ddot{R}'}\over{R'}}-2{\ddot{R}\over{R}}
+{{f-{\dot{R}^2}}\over{R^2}}
+{{(f-{\dot{R}^2})'}\over{RR'}}=-\left(1+\eta\right)
{Q_\eta}{R^{2\eta-2}}-2\Lambda = \kappa_5^2 p_z,
\eeq
is an identity for all functions $R$ and $f$. Thus we conclude that the metric has the RS-LTB form (\ref{vacs1})
and (\ref{vacs2}) with the physical radius given by Eq. (\ref{gdr2}) for $\eta\not=-1/2$ and by Eq. (\ref{gdr22})
for $\eta=-1/2$.

\subsection{Static Limits}

Of the dynamical dark radiation models the marginally bound correspond to $f=0$ and are actually static solutions.
Consider first $\eta\not=-1/2$. Transforming from the LTB coordinates $(t,r)$ to the curvature coordinates $(T,R)$
defined by
\beq
T=t+\int dR
{{\sqrt{{\Lambda\over{3}}{R^2}+{{Q_\eta}\over{2\eta+1}}
{R^{2\eta}}}}\over{
{\Lambda\over{3}}{R^2}-1+{{Q_\eta}\over{2\eta+1}}{R^{2\eta}}}}
\eeq
we find new black string solutions given by Eq. (\ref{vacs1}) with
\beq
d{s_4^2}=-\left(1-{{Q_\eta}\over{2\eta+1}}{R^{2\eta}}
-{\Lambda\over{3}}{R^2}\right)d{T^2}+
{{\left(1-{{Q_\eta}\over{2\eta+1}}{R^{2\eta}}
-{\Lambda\over{3}}{R^2}\right)}^{-1}}d{R^2}+
{R^2}d{\Omega_2^2}.\label{RNBH}
\eeq
The 4-dimensional solution (\ref{RNBH}) for $\eta=-1$ is the inhomogeneous static exterior of a collapsing sphere of
homogeneous standard dark radiation \cite{BGM,GD,RC}. When $\Lambda=0$ it corresponds to the zero mass limit of the
tidal Reissner-Nordstr\"om black hole on the brane \cite{DMPR}. The horizons covering the physical singularity at
${R_s}=0$ are defined by the transcendental equation
\beq
1-{{Q_\eta}\over{2\eta+1}}{R^{2\eta}}
-{\Lambda\over{3}}{R^2}=0.
\eeq
For $\Lambda=0$ and if the other parameters allowed it there may be an horizon located at
\beq
{R_h}={{\left({{2\eta+1}\over{Q_\eta}}\right)}^{1\over{2\eta}}}.
\eeq
If $\Lambda\not=0$ then in general it is not possible to obtain the exact location of the horizons. The two single
exceptions are the models corresponding to $\eta=-1$ and $\eta=1/2$. For $\eta=-1$ we find the standard dark radiation
horizons \cite{BGM,RC}. For $\eta=1/2$ the horizons are given by
\beq
{R_h}={{3{Q_{\mbox{\tiny
          1/2}}}}\over{4\Lambda}}\left(-1\pm\sqrt{1+{{16\Lambda}
\over{9{Q_{\mbox{\tiny 1/2}}^2}}}}\right).
\eeq
If $\Lambda<0$ and ${Q_{\mbox{\tiny 1/2}}}>0$ then we have an inner horizon $R_h^-$ and an outer horizon $R_h^+$. The
two horizons merge for ${Q_{\mbox{\tiny 1/2}}}=4\sqrt{-\Lambda}/3$ and for ${Q_{\mbox{\tiny 1/2}}}<4\sqrt{-\Lambda}/3$
the singularity at ${R_s}=0$ becomes naked. If $\Lambda>0$ and ${Q_{\mbox{\tiny 1/2}}}>0$ there is a single horizon
at $R={R_h^+}$ and for $\Lambda>0$ and ${Q_{\mbox{\tiny 1/2}}}<0$ the horizon is at $R_h^-$. Note that for $\eta=-1$
the dark radiation with a positive tidal charge satisfies all energy conditions. For $\eta=1/2$ this is no longer
true. Indeed, the strong condition is violated in 4 dimensions and in 5 dimensions none of the energy conditions holds.

For $\eta=-1/2$ we proceed analogously to find a black string given by Eq. (\ref{vacs1}) and
\bea
d{s_4^2}&=&-\left[1-{{Q_{\mbox{\tiny -1/2}}}\over{R}}\left(1+\ln R\right)
-{\Lambda\over{3}}{R^2}\right]d{T^2}+
{{\left[1-{{Q_{\mbox{\tiny -1/2}}}\over{R}}\left(1+\ln R\right)
-{\Lambda\over{3}}{R^2}\right]}^{-1}}d{R^2}\nn\\
&+&{R^2}d{\Omega_2^2}.
\eea

\subsection{Exact Dynamical Solutions}
Let us now consider the non-marginally bound models corresponding to $f\not=0$. These are the ones which actually
lead to dynamical and inhomogeneous evolutions. In general it is not possible to determine the exact solutions of
Eqs. (\ref{gdr2}) and (\ref{gdr22}). The only exceptions are $\eta=-1$ and $\eta=1/2$. If $\eta=-1$ we have
standard dark radiation and the solutions have already been determined in \cite{RC}. They are inhomogeneous
cosmologies characterized by $\Lambda$, ${Q_{\mbox{\tiny -1}}}\equiv Q$ and $f$. The corresponding rich structure
of physical singularities and regular rebounces was also identified in \cite{RC}. However, note that for
the 5-dimensional solutions defined by Eqs. (\ref{vacs1}), (\ref{vacs2}) and (\ref{gdr2}) the gravitational field
is always localized near the brane and the inhomogeneous dark
radiation dynamics has been generated on the brane by
the bulk fields. They are not a pure vaccum phenomena even if $\eta=-1$ and for such a vaccum gravity is not always
confined to the vicinity of the brane \cite{RC}.

The dynamics for $\eta=1/2$ is actually similar to that of standard
dark radiation \cite{RC}. Indeed, the solutions may also
be organized by $\Lambda$ and by the functions $Y$ and $\beta$ now defined as
\beq
Y=R+{{3{Q_{\mbox{\tiny 1/2}}}}\over{4\Lambda}},\quad
\beta={3\over{\Lambda}}\left({{3{Q_{\mbox{\tiny
            1/2}}^2}}\over{16\Lambda}}
-f\right).
\eeq
To ilustrate consider $\Lambda>0$ \cite{EXP} and allow $Q_{\mbox{\tiny
    1/2}}$ to be a real parameter as $Q$ \cite{qexp}. If $\beta>0$ then $-1<f<3{Q_{\mbox{\tiny 1/2}}^2}/(16\Lambda)$ and so the solutions
are
\beq
\left|R+{{3{Q_{\mbox{\tiny 1/2}}}}\over{4\Lambda}}\right|
=\sqrt{\beta}\cosh\left[\pm
\sqrt{{\Lambda\over{3}}}t+{\cosh^{-1}}\left(
{{\left|r+{{3{Q_{\mbox{\tiny 1/2}}}}\over{4\Lambda}}\right|}
\over{\sqrt{\beta}}}\right)\right].\label{drsol1}
\eeq
If $\beta<0$ then $f>3{Q_{\mbox{\tiny 1/2}}^2}/(16\Lambda)$ and we obtain
\beq
R+{{3{Q_{\mbox{\tiny 1/2}}}}\over{4\Lambda}}=\sqrt{-\beta}\sinh\left[\pm
\sqrt{{\Lambda\over{3}}}t+{\sinh^{-1}}
\left({{r+{{3{Q_{\mbox{\tiny 1/2}}}}\over{4\Lambda}}}
\over{\sqrt{-\beta}}}\right)\right].\label{drsol2}
\eeq
If $\beta=0$ then $f=3{Q_{\mbox{\tiny 1/2}}^2}/(16\Lambda)$ and we get
an homogeneous solution
\beq
\left|R+{{3{Q_{\mbox{\tiny 1/2}}}}\over{4\Lambda}}\right|=
\left|r+{{3{Q_{\mbox{\tiny 1/2}}}}\over{4\Lambda}}
\right|\exp\left(\pm \sqrt{{\Lambda\over{3}}}t\right).\label{drhsol}
\eeq
Clearly, solutions (\ref{drsol1}) and
(\ref{drsol2}) are intrinsically dependent on $r$ and so correspond to
inhomogeneous cosmologies which cannot be reduced to the standard
homogeneous dS or Robertson-Walker spaces \cite{HE} by any coordinate
transformation.

\subsection{Singularities and Regular Rebounces}

To analyze the space of solutions of the dark radiation models let us first take $\eta\not=-1/2$. Then we
consider Eq. (\ref{gdr1}) written as
\beq
{R^\sigma}{\dot{R}^2}=V,
\eeq
where the potential $V$ is
\beq
V=V(R,r)={{Q_\eta}\over{2\eta+1}}{R^{2\eta+\sigma}}
+{\Lambda\over{3}}{R^{2+\sigma}}
+f{R^\sigma},
\eeq
and the parameter $\sigma\geq 0$ is only non-zero when $\eta$ is negative. For example if $\eta=-1$ then
$\sigma=-2\eta=2$. Again in general it is not possible to study this potential exactly. The same is true for
$\eta=-1/2$ which involves a logarithm of $R$. Only for $\eta=-1$ and $\eta=1/2$ is such exact analysis
possible. For $\eta=-1$ a rich structure of singularities and regular rebounces was found and discussed in
\cite{RC}. A similar structure of solutions may now be shown to exist for $\eta=1/2$. To ilustrate consider
$\Lambda>0$. The potential is written as
\beq
V={\Lambda\over{3}}{R^2}+{{Q_{\mbox{\tiny 1/2}}}\over{2}}R+f=
{{\Lambda}\over{3}}\left({Y^2}-\beta\right).
\eeq
Then as for $\eta=-1$ there are at most two regular rebounce epochs and a phase of continuous accelerated
expansion to infinity.

For $\beta<0$ it is clear that $V>0$ for all values of $R\geq 0$. It satisfies $V(0,r)=f$ with
$f>3{Q_{\mbox{\tiny 1/2}}^2}/(16\Lambda)$ and it grows to infinity with $R$ as $\Lambda{R^2}$. The dark
radiation shells may either expand forever or collapse to the singularity after a proper time $t={t_s}(r)$
given by
\beq
{t_s}(r)=\sqrt{{3\over{\Lambda}}}\left[{\sinh^{-1}}
\left({{r+{{3{Q_{\mbox{\tiny 1/2}}}}\over{4\Lambda}}}
\over{\sqrt{-\beta}}}\right)-{\sinh^{-1}}\left({{3{Q_{\mbox{\tiny 1/2}}}}
\over{4\Lambda\sqrt{-\beta}}}\right)\right].
\eeq
For $\beta>0$ and independently of $Q_{\mbox{\tiny 1/2}}$ the configuration $-1<f<0$ leads to globally regular
solutions with a single rebounce epoch at $R={R_*}$ where
\beq
{R_*}=-{{3{Q_{\mbox{\tiny 1/2}}}}\over{4\Lambda}}+\sqrt{\beta}.
\eeq
Indeed, this is the minimum radius a collapsing dark radiation shell can have and it is reached after the time
$t={t_*}(r)$ where
\beq
{t_*}(r)=\sqrt{{3\over{\Lambda}}}{\cosh^{-1}}\left(
{{r+{{3{Q_{\mbox{\tiny 1/2}}}}\over{4\Lambda}}}
\over{\sqrt{\beta}}}\right).
\eeq
At this point the shells reverse their motion and expand continuously with ever increasing speed.

If $\beta>0$ and $0<f<3{Q_{\mbox{\tiny 1/2}}^2}/(16\Lambda)$ then for ${Q_{\mbox{\tiny 1/2}}}>0$ there are no
rebounce points in the allowed dynamical region $R\geq 0$. The time to reach the singularity is now given by
\beq
{t_s}(r)=\sqrt{{3\over{\Lambda}}}\left[{\cosh^{-1}}
\left({{r+{{3{Q_{\mbox{\tiny 1/2}}}}\over{4\Lambda}}}
\over{\sqrt{\beta}}}\right)-{\cosh^{-1}}\left({{3{Q_{\mbox{\tiny 1/2}}}}
\over{4\Lambda\sqrt{\beta}}}\right)\right].
\eeq
On the other hand for ${Q_{\mbox{\tiny 1/2}}}<0$ there are two rebounce epochs at $R=R_{*\pm}$ with
\beq
{R_{*\pm}}=-{{3{Q_{\mbox{\tiny 1/2}}}}\over{4\Lambda}}\pm\sqrt{\beta}.
\eeq
Since $V(0,r)=f>0$ a singularity also forms at ${R_s}=0$. The phase space of allowed dynamics is divided in two
disconnected regions separated by the forbidden interval ${R_{*-}}<R<{R_{*+}}$ where the potential is negative.
For $0\leq R\leq{R_{*-}}$ the dark radiation shells may expand to a maximum radius $R={R_{*-}}$ in the time
${t_{*-}}={t_*}$ where
\beq
{t_*}(r)=\sqrt{{3\over{\Lambda}}}{\cosh^{-1}}\left(
{{\left|r+{{3{Q_{\mbox{\tiny 1/2}}}}\over{4\Lambda}}\right|}
\over{\sqrt{\beta}}}\right).
\eeq
At this rebounce epoch the shells start to fall towards the singularity which is reached after the time
\beq
{t_s}(r)=\sqrt{{3\over{\Lambda}}}{\cosh^{-1}}
\left({{|3{Q_{\mbox{\tiny 1/2}}}|}\over{4\Lambda\sqrt{\beta}}}\right).
\eeq
If $R\geq{R_{*+}}$ then there is a collapsing phase to the minimum radius $R={R_{*+}}$ taking the time
${t_{*+}}={t_*}$ followed by reversal and subsequent accelerated continuous expansion. The singularity at ${R_s}=0$
does not form and so the solutions are globally regular.

If $\beta=0$ then $f=3{Q_{\mbox{\tiny 1/2}}^2}/(16\Lambda)$. For ${Q_{\mbox{\tiny 1/2}}}<0$ there is one rebounce
point at
\beq
{R_*}=-{{3{Q_{\mbox{\tiny 1/2}}}}\over{4\Lambda}}.
\eeq
In this case $V(0,r)=f>0$ and then a singularity also forms at ${R_s}=0$. There is no forbidden region in phase
space but the point at $R_*$ turns out to be a regular fixed point which divides two distinct dynamical regions.
Indeed if a shell starts at $R={R_*}$ then it will not move for all times. If initially $R<{R_*}$ then either
the shell expands towards $R_*$ or it collapses to the singularity. The time to get to the singularity is finite,
\beq
{t_s}(r)={3\over{\Lambda}}\ln\left({{{|3{Q_{\mbox{\tiny 1/2}}}|}\over{4\Lambda}}
\over{\left|
r+{{3{Q_{\mbox{\tiny 1/2}}}}\over{4\Lambda}}\right|}}\right),
\eeq
but the time to expand to ${R_*}$ is infinite. If initially $R>{R_*}$ then the collapsing dark radiation shells
also take an infinite time to reach $R_*$. If ${Q_{\mbox{\tiny 1/2}}}>0$ there are no real rebounce epochs and the
collapsing dark radiation simply falls to the singularity at ${R_s}=0$. The colision proper time is
\beq
{t_s}(r)=-{3\over{\Lambda}}\ln\left({{{3{Q_{\mbox{\tiny 1/2}}}}\over{4\Lambda}}
\over{r+{{3{Q_{\mbox{\tiny 1/2}}}}\over{4\Lambda}}}}\right).
\eeq

\section{Gauss-Codazzi Equations and the Localization of Gravity on the Brane }

In the special conformal setting we have defined the RS warp solution (\ref{RSwf1}) has been factored out of the
5-dimensional problem. This is then reduced to the resolution of Eqs. (\ref{4Dceq}) and (\ref{4DEeq}) with the
fifth dimensional pressure, $p_z$, satisfying conditions (\ref{eqst3}) and (\ref{5DEeqz}). A set of effective
4-dimensional brane world geometries are generated and two examples are the inhomogenous dust and dark radiation
dynamics. These effective 4-dimensional metrics should be deduced by an observer confined to
the brane which makes the same assumptions about the bulk degrees of freedom. Moreover, in this case the
4-dimensional observer should also agree about the localization of gravity in the vicinity of the brane. Let us
now show that the 5-dimensional approach developed in this work is in this sense consistent with the effective
Gauss-Codazzi formulation \cite{CGC}-\cite{RM}.

Consider then the Gauss-Codazzi approach for the RS brane world scenario and assume that
the matter degrees of freedom which exist on the brane only originate in matter modes present in the AdS bulk
space. Then the effective 4-dimensional Einstein equations are given by
\bea
{\mathcal{G}_\mu^\nu}&=&{{2{\kappa_5^2}}\over{3}}
\left[{\mathcal{T}_\alpha^\beta}
{q_\mu^\alpha}{q_\beta^\nu}+\left({\mathcal{T}_\alpha^\beta}{n^\alpha}
{n_\beta}-{1\over{4}}{\mathcal{T}_\alpha^\alpha}\right){q_\mu^\nu}
\right]+{\mathcal{K}_\alpha^\alpha}{\mathcal{K}_\mu^\nu}\nn\\
&-&{\mathcal{K}_\mu^\alpha}{\mathcal{K}_\alpha^\nu}-{1\over{2}}{q_\mu^\nu}
\left({\mathcal{K}^2}-{\mathcal{K}_\alpha^\beta}{\mathcal{K}_\beta^\alpha}
\right)
-{\mathcal{E}_\mu^\nu},\label{GCEeq1}
\eea
where
${\mathcal{G}_\mu^\nu}={G_\alpha^\beta}{q^\alpha_\mu}{q_\beta^\nu}$,
${n^\mu}={\delta_z^\mu}$ is the unit normal to the brane,
${q_\mu^\nu}={\delta_\mu^\nu}-{n_\mu}{n^\nu}$ is the tensor which
projects orthogonaly to $n^\mu$,
\beq
{\mathcal{T}_\mu^\nu}=-{\Lambda_B}{\delta_\mu^\nu}+{T_\mu^\nu}\label{GCst}
\eeq
is the stress-energy tensor,
\beq
{\mathcal{K}_\mu^\nu}={\lim_{z\to{z_0}+}}{K_\mu^\nu}=
-{{{\kappa_5^2}\lambda}\over{6}}{q_\mu^\nu}\label{GCec}
\eeq
is the extrinsic curvature and
\beq
{\mathcal{E}_\mu^\nu}={\lim_{z\to{z_0}+}}{C_{\rho\alpha\sigma\beta}}{n^\alpha}
{n^\beta}{q_\mu^\rho}{q^{\sigma\nu}}\label{weyldef}
\eeq
is the traceless projection of the 5-dimensional Weyl tensor.

Substituting Eqs. (\ref{GCst}) and (\ref{GCec}) in Eq. (\ref{GCEeq1}) we find
\beq
{\mathcal{G}_\mu^\nu}=-{{\kappa_5^2}\over{2}}\left({\Lambda_B}+
{{{\kappa_5^2}{\lambda^2}}\over{6}}\right){\delta_\mu^\nu}+
{{2{\kappa_5^2}}\over{3}}
\left[{T_\alpha^\beta}
{q_\mu^\alpha}{q_\beta^\nu}+\left({T_\alpha^\beta}{n^\alpha}{n_\beta}-
{1\over{4}}{T_\alpha^\alpha}\right){q_\mu^\nu}
\right]-{\mathcal{E}_\mu^\nu}.\label{GCEeq2}
\eeq
Applying the covariant derivative it is clear that in general the projected Weyl tensor is not conserved
because of the fields present in the bulk.

In the effective 4-dimensional point of view the metric $g_{\mu\nu}$ is defined by Eqs. (\ref{gm2}) and
(\ref{gm3}). Then we obtain ${\mathcal{E}_z^\mu}=0$. If in addition the RS identity (\ref{RSwf2}) is assumed
to be satisfied then Eq. (\ref{GCEeq2}) is written as
\beq
{G_a^b}={{2{\kappa_5^2}}\over{3}}
\left({T_a^b}+{1\over{4}}{T_z^z}{\delta_a^b}
\right)-{\mathcal{E}_a^b}.\label{GCEeq3}
\eeq
Moreover, if $T_a^b$ is conserved as in Eq. (\ref{4Dceq}) then the projected Weyl tensor must satisfy
\beq
{\nabla_a}{\mathcal{E}_b^a}={{\kappa_5^2}\over{6}}{\nabla_b}{T_z^z}.
\eeq
If it is verified that
\beq
{\mathcal{E}_a^b}={{\kappa_5^2}\over{3}}
\left(-{T_a^b}+{1\over{2}}{T_z^z}{\delta_a^b}
\right)\label{GCweyl1}
\eeq
then Eq. (\ref{GCEeq3}) becomes Eq. (\ref{4DEeq}) and so the 4-dimensional observer does find the same
dynamics on the brane.

This may be explicitly checked for the dust and dark radiation systems. First determine $\mathcal{E}_\mu^\nu$
using the alternative Eqs. (\ref{weyldef}) and (\ref{GCweyl1}). For dust both lead to the same result
\beq
{\mathcal{E}_t^t}={{\kappa_5^2}\over{4}}{\rho_{\mbox{\tiny D}}},\quad
{\mathcal{E}_r^r}={\mathcal{E}_\theta^\theta}={\mathcal{E}_\phi^\phi}=
-{{\kappa_5^2}\over{12}}{\rho_{\mbox{\tiny D}}}.\label{dustE}
\eeq
Thus Eq. (\ref{GCweyl1}) is indeed verified and then it is easy to see that Eq. (\ref{GCEeq3}) reduces to the
4-dimensional Einstein equations for dust and a cosmological constant
\beq
{G_t^t}=-{\kappa_5^2}{\rho_{\mbox{\tiny D}}}-\Lambda,\quad
{G_r^r}={G_\theta^\theta}={G_\phi^\phi}=
-\Lambda.
\eeq
For the dark radiation system we also conclude that both Eqs. (\ref{weyldef}) and (\ref{GCweyl1}) lead to the
same result
\beq
{\mathcal{E}_t^t}={\mathcal{E}_r^r}=
{{\kappa_5^2}\over{6}}(1-\eta){\rho_{\mbox{\tiny DR}}},\quad
{\mathcal{E}_\theta^\theta}={\mathcal{E}_\phi^\phi}=-{{\kappa_5^2}\over{6}}
(1-\eta){\rho_{\mbox{\tiny DR}}}\label{drE}
\eeq
so that Eq. (\ref{GCweyl1}) is satisfied. Then Eq. (\ref{GCEeq3}) reduces to the 4-dimensional Einstein equations
for dark radiation and a cosmological constant
\beq
{G_t^t}={G_r^r}=-{\kappa_5^2}{\rho_{\mbox{\tiny
      DR}}}
-\Lambda,\quad{G_\theta^\theta}={G_\phi^\phi}={\kappa_5^2}
{\rho_{\mbox{\tiny
      DR}}}
-\Lambda.
\eeq
In the 5-dimensional picture it is clear that whatever the effective 4-dimensional solution to be considered,
the conformal RS warp factor ensures that gravity is always localized in the vicinity of the brane. Let us now
show that the observer confined to the brane may reach the same conclusion. In the covariant Gauss-Codazzi
approach the tidal acceleration away from the brane is defined by
\cite{RM,RM4dp}
\beq
{a_T}=-{\lim_{z\to{z_0}+}}{R_{\mu\nu\alpha\beta}}{n^\nu}{u^\nu}
{n^\alpha}{n^\beta},
\eeq
where ${u^\mu}={\delta_t^\mu}$ is the extension off the brane of the 4-velocity field which satisfies ${u^\mu}{n_\mu}=0$
and ${u^\mu}{u_\mu}=-1$. For the gravitational field to be localized near the brane $a_T$ must be negative. Consider
the identity \cite{SMS}
\beq
{R_{\mu\nu\alpha\beta}}={C_{\mu\nu\alpha\beta}}+{2\over{3}}\left\{
{g_{\mu[\alpha}}{R_{\beta]\nu}}+{g_{\nu[\beta}}{R_{\alpha]\mu}}\right\}-
{1\over{6}}R{g_{\mu[\alpha}}{g_{\beta]\nu}},
\eeq
where the square brackets denote anti-symmetrization. Introducing the 5-dimensional Einstein equation
\beq
{G_\mu^\nu}={\kappa_5^2}{\mathcal{T}_\mu^\nu}
\eeq
we find for the tidal acceleration the following expression
\beq
{a_T}={{{\kappa_5^2}{\Lambda_B}}\over{6}}-{{\kappa_5^2}\over{3}}\left(
{T_\alpha^\beta}{u^\alpha}{u_\beta}-{T_z^z}+{{T_\alpha^\alpha}\over{2}}
\right)-
{\mathcal{E}_\alpha^\beta}{u^\alpha}{u_\beta}.
\eeq
If the contributions of $T_\mu^\nu$ and $\mathcal{E}_\mu^\nu$ cancel each other,
\beq
{\mathcal{E}_\alpha^\beta}{u^\alpha}{u_\beta}=-{{\kappa_5^2}\over{3}}\left(
{T_\alpha^\beta}{u^\alpha}{u_\beta}-{T_z^z}+{{T_\alpha^\alpha}\over{2}}\right)
,\label{GCweyl2}
\eeq
then the tidal acceleration is given by
\beq
{a_T}={{{\kappa_5^2}{\Lambda_B}}\over{6}},\label{GCta}
\eeq
which is indeed always negative in the RS 5-dimensional AdS space.

This may also be explicitly confirmed for the dust and dark radiation systems. For dust Eqs. (\ref{dustes1})
and (\ref{dustes2}) imply
\beq
{a_T}={{{\kappa_5^2}{\Lambda_B}}\over{6}}-{{\kappa_5^2}\over{4}}{\rho_D}-
{\mathcal{E}_\alpha^\beta}{u^\alpha}{u_\beta}.
\eeq
Then using Eq. (\ref{dustE}) it is easy to see that Eq. (\ref{GCweyl2}) is indeed verified. In the vaccum
the solutions are conformally flat and so we also obtain the same result.
Note, however, that using the effective Gauss-Codazzi approach it is possible to find a much richer non-conformally
flat vaccum dynamics which also admits a brane cosmological constant \cite{RC}. Then it turns out that gravity
is not always confined to the vicinity of the brane \cite{RC}. On the other hand for dark radiation generated on
the brane by the conformal bulk fields Eqs. (\ref{gdres1})-(\ref{gdres3}) imply
\beq
{a_T}={{{\kappa_5^2}{\Lambda_B}}\over{6}}-{{\kappa_5^2}\over{3}}(1-\eta)
{\rho_{\mbox{\tiny
      DR}}}-
{\mathcal{E}_\alpha^\beta}{u^\alpha}{u_\beta}.
\eeq
Once more Eq. (\ref{GCweyl2}) is verified as may be checked introducing Eq. (\ref{drE}). Then the gravitational
field is always bound to the vicinity of the brane.

\section{Conclusions}

Any theory in which our universe is viewed as a brane must reproduce the large scale
predictions of 4-dimensional general relativity, in particular the gravitational
collapse of matter, on it. In the bulk, this matter could be localized about the brane
or extended.  In this work, we have analyzed the dynamical collapse of extended matter in a
$Z_2$ symmetric, 5-dimensional world in which a single spherically symmetric,
positive tension, gravity confining brane exists. We have
applied a global conformal transformation to give a clarifying
organization to the Einstein equations for the
most general 5-dimensional metric consistent with the $Z_2$ symmetry
and with the spherical symmetry on the brane.
Assuming that the bulk stress-energy tensor has conformal
weight $s=-4$ and that only the conformal warp factor depends on the
coordinate of the fifth dimension, we showed that the bulk matter dynamics on the brane
produces a pressure along the fifth dimension which is
required to satisfy a well defined equation of
state.

With this analysis we have discovered a new class of exact dynamical
solutions for which the conformal warp factor localizes gravity in the
vicinity of the brane. We have considered the two specific examples of
5-dimensional geometries which define on the brane
the gravitational dynamics of inhomogeneous dust and generalized
dark radiation in the presence of an effective brane cosmological constant,
also generated by bulk matter. For these examples we have discussed the static
marginally bound limits and the conditions defining the solutions as singular
or as globally regular.

Finally, we have also analyzed the point of view of an observer
confined to the brane to show that an identical
description of the variables in the problem consistently leads to the
same localized brane world dynamics.

The bulk matter we consider is extended and not localized near the brane
because the conformal weight of the stress-energy tensor is $s=-4$. Its
density and pressures depend on the coordinate of the fifth dimension and
diverge on the horizon. This problem is also encountered by the black string
solution \cite{CHR}, where the Kretschmann invariant is shown to diverge on
the AdS horizon. It can be solved if we use the first RS model in which
there are two branes with opposite tensions present and we live on the
brane with negative tension.  Alternatively, one can look for solutions in
a single brane model that respect the localization of gravity
while incorporating a simultaneous localization of matter about the brane.
This would amount to looking for solutions in which the conformal metric is
$g_{\mu\nu}(t,r,z)$, with $g_{\mu\nu}(t,r,z=z_0) = g_{\mu\nu}(t,r)$ where
$g_{\mu\nu}(t,r)$ corresponds to one of the class of solutions we have presented
plus (possibly) quantum corrections \cite{EFK} (see \cite{KT} for a similar
approach regarding static black hole solutions). This will be examined in
future work.
\vspace{1cm}

\centerline{\bf Acknowledgements}
\vspace{0.25cm}

We are grateful for financial support from the {\it Funda\c {c}\~ao para a
Ci\^encia e a Tecnologia} (FCT) as well as the {\it Fundo Social Europeu}
(FSE) under the contracts SFRH/BPD/7182/2001 and POCTI/32694/FIS/2000
({\it III Quadro Comunit\'ario de Apoio}).

\end{document}